\PassOptionsToPackage{table}{xcolor}
\documentclass{article} 
\usepackage{iclr2025_conference,times}

\usepackage{amsmath,amsfonts,bm}

\def\eqref#1{equation~\ref{#1}}

\def\1{\bm{1}}

\DeclareMathAlphabet{\mathsfit}{\encodingdefault}{\sfdefault}{m}{sl}
\SetMathAlphabet{\mathsfit}{bold}{\encodingdefault}{\sfdefault}{bx}{n}

\usepackage{hyperref}
\usepackage{url}

\usepackage{graphicx}
\usepackage{multirow}    
\usepackage{booktabs}    
\usepackage{colortbl}    
\usepackage[table]{xcolor} 
\usepackage{subcaption}

\NewDocumentCommand\emojititle{}{
    $\vcenter{\hbox{\includegraphics[height=1.2em]{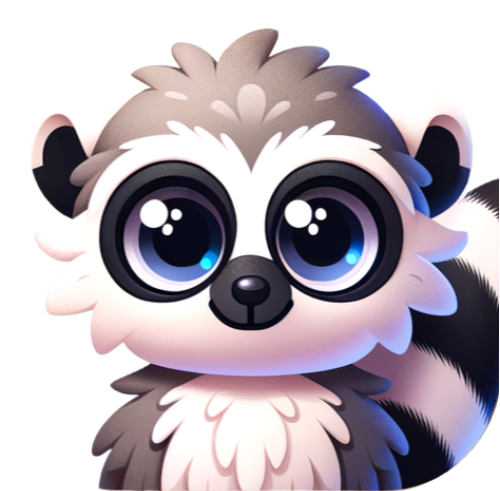}}}$
}
\newcommand{\model}{\textsc{Lemur}}

\title{\emojititle{}\model{}: Log Parsing with Entropy Sampling and Chain-of-Thought Merging}

\author{Wei Zhang$^{1}$, Xiangyuan Guan$^{1}$, Lu Yunhong$^{2}$, Jie Zhang$^{3}$, \\\textbf{Shuangyong Song}$^{3}$, \textbf{Xianfu Cheng}$^{1}$, \textbf{Zhenhe Wu}$^{1}$, \textbf{Zhoujun Li}$^{1}$
\\\textsuperscript{\rm 1}State Key Lab of Software Development Environment, Beihang University
\\\textsuperscript{\rm 2}Yantai University
\\\textsuperscript{\rm 3}China Telecom
\\\texttt{zwpride@buaa.edu.cn}
}

\iclrfinalcopy 
\begin{document}

\maketitle

\begin{abstract}
Logs produced by extensive software systems are integral to monitoring system behaviors. Advanced log analysis facilitates the detection, alerting, and diagnosis of system faults. Log parsing, which entails transforming raw log messages into structured templates, constitutes a critical phase in the automation of log analytics. Existing log parsers fail to identify the correct templates due to reliance on human-made rules. Besides, these methods focus on statistical features while ignoring semantic information in log messages. To address these challenges, we introduce a cutting-edge \textbf{L}og parsing framework with \textbf{E}ntropy sampling and chain-of-thought \textbf{M}erging (\model{}). Specifically, to discard the tedious manual rules, we propose a novel sampling method inspired by information entropy, which efficiently clusters typical logs. Furthermore, to enhance the merging of log templates, we design a chain-of-thought method for large language models (LLMs). LLMs exhibit exceptional semantic comprehension and deftly distinguish between parameters and invariant tokens. We have conducted experiments on large-scale public datasets. Extensive evaluation demonstrates that \model{} achieves state-of-the-art performance and impressive efficiency. The Code is available at \url{https://github.com/zwpride/lemur}.
\end{abstract}

\section{Introduction}

\begin{figure}[t]
\centering
\includegraphics[width=0.7\linewidth]{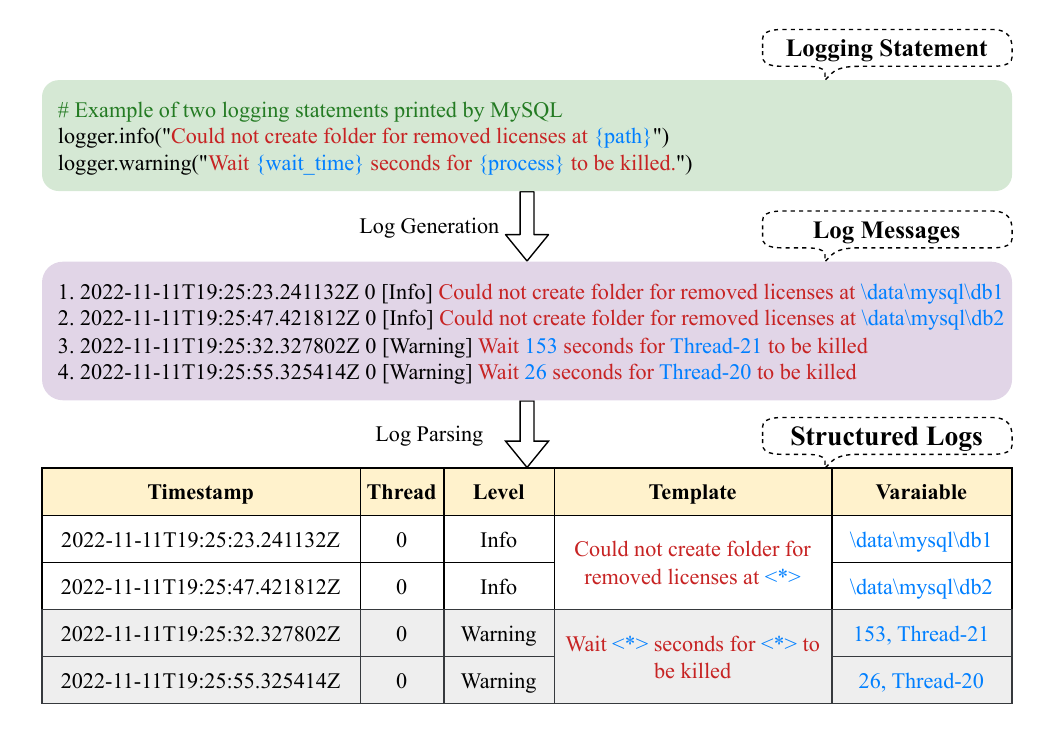}
\caption{An example of log parsing. \textbf{Logging Statement} cannot be accessed in most cases. \textbf{Log Message} is parsed into a static \textbf{Template} part containing fixed keywords and a \textbf{Variable} part that will vary between different log messages with the same template.}
\label{fig1}
\vspace{-15pt}
\end{figure}

Logs serve as a critical information  for system monitoring, offering key insights into system behavior, a fact well-documented in existing literature \citep{automated_log_analysis}. Their advantage over other types of data lies in their rich informational content and relative ease of interpretation. Through log analysis, several important downstream tasks can be effectively addressed, which include anomaly detection \citep{deeplog,cfg_mining}, fault diagnosis \citep{fault_identifying,fault_uilog}, and root cause analysis~\citep{root_logan}. Log parsing, a crucial initial step in log analysis \citep{he2016evaluation}, separates log messages into two parts: 1) Log Templates. The constant, unchanging parts in logging statements, and 2) Log Variables. The dynamic, changeable details in different executions. In Figure~\ref{fig1}, the logging statement \texttt{logger.info('Wait \{wait\_time\} seconds for \{process\} to be killed.')} can yield various messages like \texttt{Wait 26 seconds for Thread-20 to be killed}. \texttt{Wait $<*>$ seconds for $<*>$ to be killed} is log template, and the changing data like \texttt{26} and \texttt{Thread-20} are log variables.

In the field of system maintenance, where source code accessibility is often restricted, different parsers \citep{drain, drain+, logppt, brain,llmparser} have been developed to differentiate between templates and variables within log messages autonomously. Existing log parsers can be categorized into two groups: syntax-based and semantic-based. Syntax-based log parsers \citep{drain, spell, iplom, slct,logcluster} utilize specific features or heuristics (e.g., word frequency) to extract the constant parts of log messages as templates. Conversely, semantic-based parsers~\cite{uniparser,logppt} leverages advanced deep learning frameworks to assimilate semantics and system-specific patterns from annotated log data, thereby facilitating the parsing of new logs. Besides, recent works~\cite{llmparser} leverage large language models (LLMs) \citep{gpt4} for log parsing, which also utilizes the powerful semantic understanding of LLMs. 

However, syntax-based methodologies are heavily dependent on meticulously crafted rules. Their performance significantly diminishes with the exponential increase in the volume and intricacy of log data. Furthermore, these techniques often overlook semantic variances among logs. For instance, logs such as \texttt{success to open file/var/log/system} and \texttt{success to close file/var/log/system} display syntactic resemblance, potentially leading to their aggregation by data-driven approaches, which could adversely impact the efficacy of anomaly detection tasks. Semantic-based methods, reliant on pre-labeled logs for training, falter in understanding semantics when applied to unfamiliar domains. LLM-based parsers \citep{logprompt, logppt, llmparser} have lower availability due to high inference time (such as GPUs) and network latency \citep{jiao2023chatgpt}. Besides, LLMs generate unstable results because of the hallucination problem. 

To tackle these challenges, we propose \model{}, a cutting-edge \textbf{L}og parsing framework with \textbf{E}ntropy sampling and Chain-of-Thought \textbf{M}erging (\model{}). \model{} brings together the strengths of the syntax-based and semantic-based methods, which consist of three key components: Information Entropy Clustering, Template Generation, and Chain-of-Thought Merging. Specifically, inspired by information theory~\cite{1entropy}, we recognize that different logs encapsulate varying quantities of information, while logs of a similar nature contain comparable amounts. Consequently, we have developed a novel sampling method based on information entropy principles that efficiently clusters characteristic logs by dividing large-scale data into multiple clusters and utilizing efficient sampling and clustering algorithms within those clusters, thereby ensuring that LEMUR remains robust and high-performance in large-scale log scenarios. Then in template generation, we determine the variables and the template in the log based on the information entropy of the token at the same location. In the end, motivated by Chain-of-Thought~\cite{chain-of-thought},  we design a three-hop Chain-of-Thought (infer structure, infer semantics, and infer solution) for merging templates.

\begin{figure*}[t]
    \centering
    \includegraphics[width=1.0\textwidth]{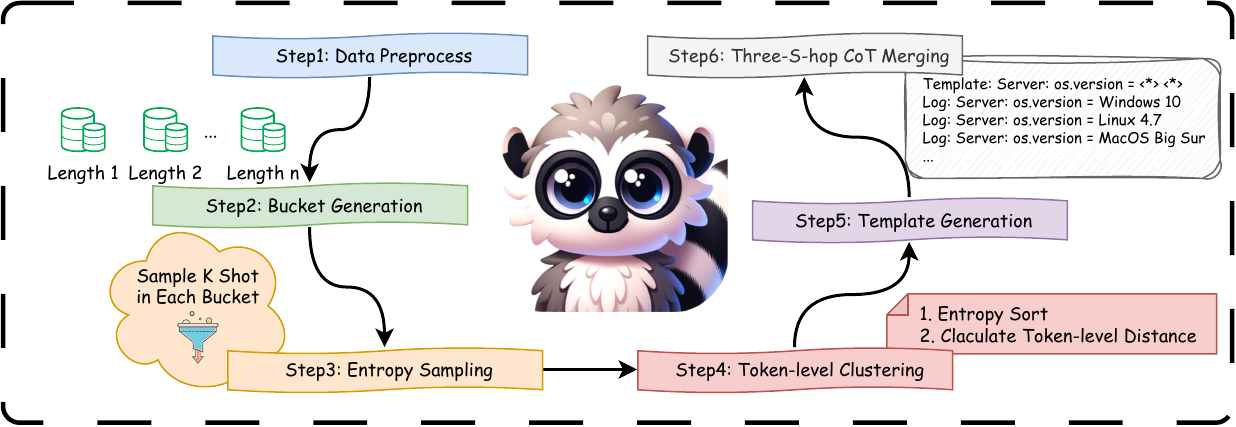}
    \caption{Overview of our log parsing framework.}
    \label{fig: framework}
    \vspace{-10pt}
\end{figure*}

We have conducted a comprehensive evaluation on public large-scale log datasets of LogHub~\cite{loghub} with seven state-of-the-art log parsing approaches. The results show that \model{} achieves the highest performance than other baselines for the F1 score of grouping and template accuracy. Generally, the main contributions of this work are listed as follows: 
\begin{itemize}
    \item To the best of our knowledge, we propose \model{}, the first unsupervised framework to combine information entropy and large language models for online log parsing.
    \item We introduce a novel sampling method based on information entropy for efficient log clustering. Besides, by utilizing LLMs, we can accurately merge templates based on semantics rather than syntax-based methods.
    \item Extensive experiments are conducted on public benchmarks to demonstrate the effectiveness of our \model{}. The results show that \model{} outperforms other state-of-the-art methods.
\end{itemize}

\section{Methodology}
In this section, we overview \model{}, a framework designed to enhance log message categorization. Figure~\ref{fig: framework} shows \model{}, which consists of three main components: Information Entropy Clustering, Template Generation, and Chain-of-Thought Merging.

\subsection{Information Entropy Clustering}
In this section, logs are categorized into buckets based on their length. Some representative logs are selected as the cluster centers within each bucket, and other logs are clustered around these centers.

\paragraph{Bucket Generation}
Logs are assigned to buckets $B_j$ by a mapping function $f$ based on length: $f: L \rightarrow B$, with $L$ representing logs and $B$ representing buckets. For each log $L_i \in L$, we compute its length $len(L_i)$ after tokenization. The log is then assigned to bucket $B_j \in B$, where $j$ matches the index of the log's length in the set of unique lengths $l$. This can be written as $f(L_i) = B_j$ where $l_j = len(L_i)$. This method ensures logs of the same length are grouped, reducing variability and aiding subsequent analysis.

\paragraph{Entropy-based Sampling}
To identify potential log templates from buckets of logs with identical token lengths, we propose a clustering method inspired by information theory~\cite{1entropy}, specifically employing Shannon entropy. Unlike conventional clustering which uses random seeds and iterative refinement, our approach selects logs based on their information content, which is ideal for log analysis due to the variability in log messages.

Each log $x \in B_j$ is evaluated by its entropy $E(x) = -\sum_{i} p(x_i) \log p(x_i)$, where $x_i$ is the token and $p(x_i)$ its probability, gauging its information content. Logs are then ranked by descending entropy into layers ${Lay_1, \ldots, Lay_n}$ to prioritize those with rich information.

We select clustering centers from these layers, starting with the highest entropy logs and picking those with either new first tokens or entropy above a threshold. This process repeats until we've chosen $k$ centers or the remaining logs no longer offer new information. Our stopping criterion ensures we gather diverse and informative logs while avoiding redundancy.

\paragraph{Refinement with Jaccard Similarity}
After obtaining the initial set of $k$ samples, we further refine the selection by merging similar log centers based on their Jaccard similarity. This merging process helps eliminate redundancy while preserving the diversity of log samples. We use the Jaccard similarity to measure the similarity between two logs. The Jaccard similarity between two logs $L_1$ and $L_2$ is calculated as $similarity = \frac{|L_1 \cap L_2|}{|L_1 \cup L_2|}$. Let $J_T$ be the threshold for Jaccard similarity; if the similarity between two centers ($L_1$, $L_2$) exceeds $J_T$, we remove $L_2$. This refinement ensures a diverse and representative set of $k'$ log samples and optimizes the balance between coverage and conciseness in the log samples.

\paragraph{Token-level Clustering}
In the clustering process for logs within a bucket $B_j \in B$, we distinguish between the set of representative logs, $S$, which serve as cluster centers, and the set of remaining logs, $O$, which are to be clustered around these centers.

Let $S = \{s_1, s_2, \ldots, s_k\}$ represent the cluster centers and $O = \{o_1, o_2, \ldots, o_n\}$ denote the logs to be clustered. The objective is to assign each log $o_i \in O$ to a cluster center $s_j \in S$ such that the edit distance between them, $d(o_i, s_j)$, is minimized. The edit distance measures the similarity between two logs in terms of the changes required to convert one into the other at the token level. The edit distance $d(o_i, s_j)$ between a log $o_i$ and a cluster center $s_j$ is defined as:
\[
d(o_i, s_j) = \min_{s_j \in S} \sum_{t=1}^{T} \delta(o_{i,t}, s_{j,t})
\]
where $\delta(o_{i,t}, s_{j,t})$ is the token-level edit distance between the $t$-th token of $o_i$ and $s_j$, and $T$ represents the total number of tokens.

By minimizing $d(o_i, s_j)$ for each log $o_i$, we cluster logs around their nearest representative center in $S$, ensuring that logs within a cluster are as similar as possible according to the defined metric.

\begin{figure}[!t]
    \centering
    \includegraphics[width=0.9\columnwidth]{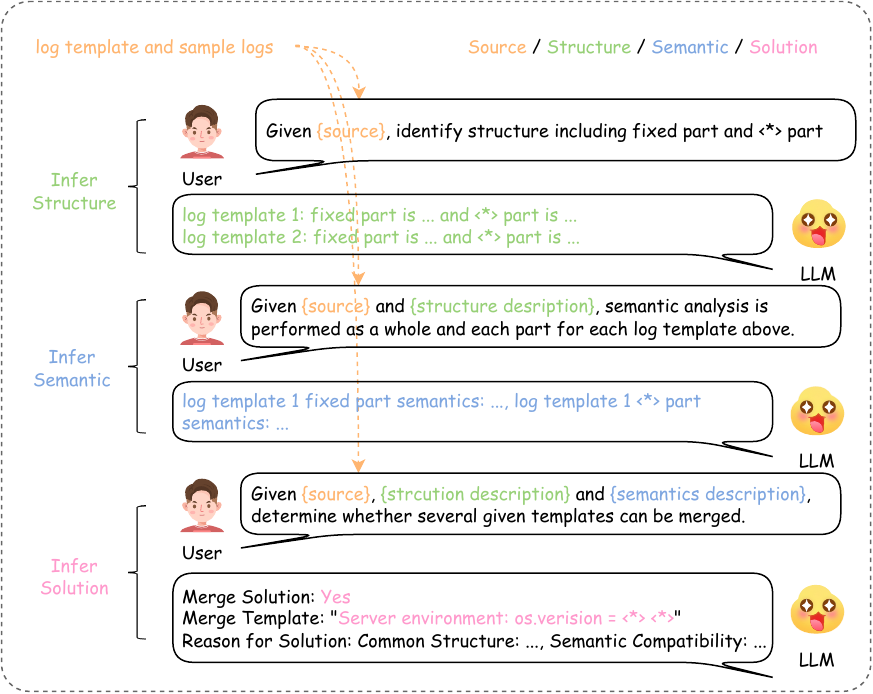}
    \caption{Three-S-hop Chain-of-Thought Merging Method}
    \label{fig:merge_log_template_prompt}
    \vspace{-20pt}
\end{figure}

\subsection{Template Generation}
Template generation aims to differentiate variables from fixed components in logs. It starts with finding the Longest Common Subsequence (LCS), which indicates shared elements across logs. Then, each log is compared with the LCS to pinpoint unique tokens and their locations; for example, comparing \texttt{ABC} with the LCS \texttt{B} reveals distinct tokens at positions \texttt{A-0} and \texttt{C-2}.

For each unique position, we calculate information entropy inline using $H(X) = -\sum P(x_i) \log_2 P(x_i)$, where $P(x_i)$ is the probability of the token at that position, and compile a list of these tokens.

To classify a position as a variable, we apply the inline decision: a position is variable if $H > \theta$, otherwise fixed, where $H$ represents information entropy and $\theta$ is a set threshold depending on data characteristics and iterative experiments. When $H$ surpasses $\theta$, we substitute all tokens at that position with the variable marker $<*>$.

Additionally, we mark digit-dominated tokens as variables $<*>$ and use NLP to identify fixed-value tokens (e.g., hash values), enhancing variable identification accuracy. This method effectively identifies variables and fixed parts in logs, adapts to different types of logs, and accurately identifies changing parts while preserving the log structure.

\subsection{Three-S-hop Chain-of-Thought Merging}

In advanced log message analysis, after categorizing messages into distinct clusters and extracting unique log templates, we encounter a significant challenge: reconciling log templates that are semantically similar but differ in length. Despite structural variations, these templates often represent identical or closely related events. This challenge primarily arises from the limitations of length-based classification methods, which frequently fail to recognize that templates with equivalent semantics but different expressions may correspond to the same event type.

Illustrated in Figure \ref{fig:merge_log_template_prompt}, We propose \textit{Three-S-hop Chain-of-Thought Merging} using Large Language Models (LLMs) for semantic parsing. This includes data processing and three dialogue rounds: Structure QA (examines structure and length differences), Semantic QA (probes meanings for semantic equivalences), and Solution QA (decides on merging based on prior analyses). This approach merges templates of different lengths but similar meanings through structural and semantic analysis, enhancing accurate identification.



\begin{table}[!t]
\begin{center}
\resizebox{\columnwidth}{!}{
\begin{tabular}{lcccccccccccccccc}
\toprule
 & HDFS & Hadoop & Spark & Zookeeper & BGL & HPC & Thunderbird & Windows & Linux & Android & HealthApp & Apache & Proxifier & OpenSSH & OpenStack & Mac \\
\midrule
Template & 14 & 114 & 36 & 50 & 120 & 46 & 149 & 50 & 118 & 166 & 75 & 6 & 8 & 27 & 43 & 341 \\
Average Length & 7.44 & 8.19 & 8.76 & 6.30 & 6.32 & 3.48 & 8.51 & 7.93 & 8.30 & 5.40 & 2.80 & 6.28 & 9.35 & 8.56 & 9.01 & 9.17 \\
Various Lengths Proportion & 0.25\% & 0.00\% & 0.00\% & 0.55\% & 0.00\% & 0.40\% & 2.25\% & 0.00\% & 0.00\% & 3.85\% & 0.00\% & 0.00\% & 47.35\% & 0.00\% & 0.00\% & 0.35\% \\
Messages & 2000 & 2000 & 2000 & 2000 & 2000 & 2000 & 2000 & 2000 & 2000 & 2000 & 2000 & 2000 & 2000 & 2000 & 2000 & 2000 \\
\bottomrule
\end{tabular}
}
\end{center}
\caption{Summary of LogHub datasets.}
\label{summary_of_public_datasets_transposed}
\end{table}

\begin{table}[!t]
\begin{center}
\resizebox{\textwidth}{!}{
\begin{tabular}{c|cccccccccccccccc}
\toprule
 & HDFS & Hadoop & Spark & Zookeeper & BGL & HPC & Thunderbird & Windows & Linux & Android & HealthApp & Apache & Proxifier & OpenSSH & OpenStack & Mac \\
\midrule
$D$ & : & =, :, (, ), \_ & : & =, :, , & =, ., ., (, ) & =, -, : & :, = & =, :, [, ] & =, : & (, ) & =, : &  &  & = &  &  \\
$k$ & 2 & 8 & 6 & 8 & 9 & 9 & 11 & 8 & 25 & 9 & 12 & 12 & 12 & 4 & 20 & 12 \\
$J_T$ & 0.7 & 0.7 & 0.6 & 0.9 & 0.6 & 0.6 & 0.4 & 0.6 & 0.33 & 0.7 & 0.7 & 0.7 & 0.7 & 0.5 & 0.7 & 0.7 \\
$\theta$ & 2.0 & 1.7 & 2.1 & 2.2 & 5.5 & 1.2 & 4.1 & 1.1 & 0.09 & 3.5 & 0 & 0 & 0.1 & 0.2 & 2.3 & 4.7 \\
\bottomrule
\end{tabular}
}
\end{center}
\label{hyperparameters_setting_of_public_datasets_transposed}
\caption{Transposed hyperparameters setting of Loghub datasets. $D$ is the tokens for word-splitting, $k$ denotes the number of the $S$ in each bucket, $J_T$ is the Jaccard similarity threshold for within-bucket merging, and $\theta$ is the entropy threshold for token identification.}
\end{table}

\section{Experiment}
\subsection{Implement Details}
\paragraph{Datasets} 
Experiments are conducted on the most widely-used benchmark datasets published in LogHub \citep{loghub}. More details are available in Table 1. 

\paragraph{Implementation and Configuration} 
We implement \model{} based on Python 3.10, Apple M3 chip with 8-core CPU, 10-core GPU 16GB RAM, and macOS Sonoma(14.2.1). In experiments, $D$ is the token set for word-splitting, $k$ denotes the number of the $S$ in each bucket, the Jaccard similarity threshold for within-bucket merging $J_T$ and the entropy threshold for token identification $\theta$ are shown in Table 2. We use \textbf{GPT-4} as the base to implement the three-S-hop chain of thought merging, and \model{} is suitable for other LLMs. \model{} uses these models to conduct inference just by locally loading the weights of these models or utilizing the APIs provided by these LLMs.

\begin{table*}[!t]
\begin{center}
\resizebox{1.0\textwidth}{!}{
 \small
\begin{tabular}{ccccccccccccccc>{\columncolor{gray!50}}c>{\columncolor{gray!50}}c}
   \toprule

   \multirow{2}*{Dataset}
   &\multicolumn{2}{c}{Drain}&\multicolumn{2}{c}{Spell}&\multicolumn{2}{c}{IPLOM}&\multicolumn{2}{c}{ULP}&\multicolumn{2}{c}{Brain}&\multicolumn{2}{c}{LogPPT}&\multicolumn{2}{c}{LLMParser}&\multicolumn{2}{c}{\model{}}\\

   \cmidrule(lr){2-3}\cmidrule(lr){4-5}\cmidrule(lr){6-7}\cmidrule(lr){8-9}\cmidrule(lr){10-11}\cmidrule(lr){12-13}\cmidrule(lr){14-15}\cmidrule(lr){16-17}
   
   &FGA&GA&FGA&GA&FGA&GA&FGA&GA&FGA&GA&FGA&GA&FGA&GA&FGA&GA\\
    
   \midrule
    HDFS&0.999&0.998&\textbf{1}&\textbf{1}&\textbf{1}&\textbf{1}&0.999&0.998&0.999&0.998&0.957&0.845&0.965&\textbf{1}& \textbf{1}&\textbf{1}\\
    Hadoop&0.999&0.948&0.920&0.777&0.996&0.954&0.999&0.950&0.999&0.949&0.999&0.977&0.958&\textbf{1}&\textbf{1}&\textbf{1}\\
    
    Spark&0.992&0.920&0.991&0.905&0.992&0.920&0.995&0.922&0.999&0.998&0.997&0.848&0.906&0.995&\textbf{1}&\textbf{1}\\
    
    Zookeeper&0.999&0.967&0.999&0.964&0.999&0.993&0.999&0.988&0.999&0.985&\textbf{1}&\textbf{1}&0.967&\textbf{1}&\textbf{1}&\textbf{1}\\
    
    BGL&0.999&0.963&0.957&0.786&0.999&0.939&0.999&0.930&0.999&0.986&0.968&0.455&0.864&0.892&\textbf{1}&\textbf{1}\\
    
    HPC&0.991&0.887&0.986&0.654&0.978&0.829&0.994&0.951&0.998&0.945&0.999&0.941&0.910&0.872&\textbf{1}&\textbf{1}\\
    
    Thunderbird&\textbf{0.999}&0.955&0.994&0.844&\textbf{0.999}&0.663&\textbf{0.999}&0.675&\textbf{0.999}&0.971&0.714&0.262&0.799&0.813&\textbf{0.999}&\textbf{0.982}\\
    
    Windows&0.999&0.997&0.999&0.989&0.995&0.567&0.989&0.410&0.999&0.997&0.992&0.717&0.984&0.783&\textbf{1}&\textbf{1}\\
    
    Linux&0.992&0.690&0.937&0.605&0.964&0.671&0.476&0.363&0.999&0.996&0.713&0.177&0.921&0.961&\textbf{0.999}&\textbf{0.988}\\
    
    Andriod&0.996&0.911&0.992&0.919&0.949&0.712&0.971&0.838&0.997&0.961&0.989&0.862&0.990&0.873&\textbf{0.999}&\textbf{0.995}\\
    
    HealthApp&0.918&0.780&0.887&0.639&0.958&0.822&0.993&0.901&\textbf{1}&\textbf{1}&\textbf{1}&0.999&0.978&\textbf{1}&\textbf{1}&\textbf{1}\\
    
    Apache&\textbf{1}&\textbf{1}&\textbf{1}&\textbf{1}&\textbf{1}&\textbf{1}&\textbf{1}&\textbf{1}&\textbf{1}&\textbf{1}&0.999&0.582&\textbf{1}&\textbf{1}&\textbf{1}&\textbf{1}\\
    Proxifier&0.785&0.526&0.832&0.526&0.786&0.516&0.940&0.024&\textbf{1}&\textbf{1}&\textbf{1}&\textbf{1}&\textbf{1}&\textbf{1}&\textbf{1}&\textbf{1}\\
    
    OpenSSH&0.999&0.787&0.918&0.554&0.998&0.540&0.940&0.434&\textbf{1}&\textbf{1}&0.983&0.436&0.836&0.697&\textbf{1}&\textbf{1}\\
    
    OpenStack&0.993&0.733&0.994&0.764&0.909&0.331&0.834&0.491&\textbf{1}&\textbf{1}&0.997&0.492&\textbf{1}&\textbf{1}&\textbf{1}&\textbf{1}\\
    Mac&0.975&0.786&0.963&0.756&0.957&0.670&0.981&0.814&0.996&0.942&0.720&0.761&0.830&0.871&\textbf{0.984}&\textbf{0.977}\\
    
    \midrule
    \textbf{Average}&0.977&0.865&0.961&0.793&0.968&0.756&0.932&0.733&0.999&0.983&0.939&0.710&0.931&0.913&\textbf{0.999}&\textbf{0.996}\\
   \bottomrule
\end{tabular}
}
\end{center}
\caption{FGA and GA on LogHub Dataset.}
\label{public_data_accuracy}
\end{table*}

\subsection{Baselines and Metrics}
\paragraph{Baselines}
As for baselines, we choose Drain \citep{drain}, Spell \citep{spell}, IPLOM \citep{iplom}, ULP \citep{ulp}, Brain \citep{brain}, LogPPT \citep{logppt} and LLMParser \citep{llmparser} as our baselines.

\paragraph{F1 score of Grouping Accuracy (FGA)}
FGA is a template-level metric that focuses on the ratio of correctly grouped templates. Specifically, let $N_g$ be the actual correct number of templates in the ground truth, and $N_p$ be the number of templates that are generated by a log parser. If $N_c$ is the number of templates that are correctly parsed by the log parser, then we can compute the Precision of Grouping Accuracy (PGA) as $\frac{N_c}{N_p}$ and the Recall of Grouping Accuracy (RGA) as $\frac{N_c}{N_g}$. The FGA is equal to their harmonic mean, \emph{ie} $\frac{2 \times GPA \times RGA}{PGA + RGA}$.

\paragraph{Grouping Accuracy (GA)}
GA is computed as the ratio of correctly grouped log messages to the total count of log messages. A log message is considered to be correctly grouped if and only if its template aligns with the same set of log messages as that of the ground truth.

\paragraph{Execution time.} We measure the execution time in seconds and compare \model{} with other parsers in terms of efficiency.

\subsection{Main Results}

\begin{figure*}[t]
    \centering
    \begin{subfigure}[t]{0.45\textwidth}
        \centering
        \includegraphics[width=\textwidth]{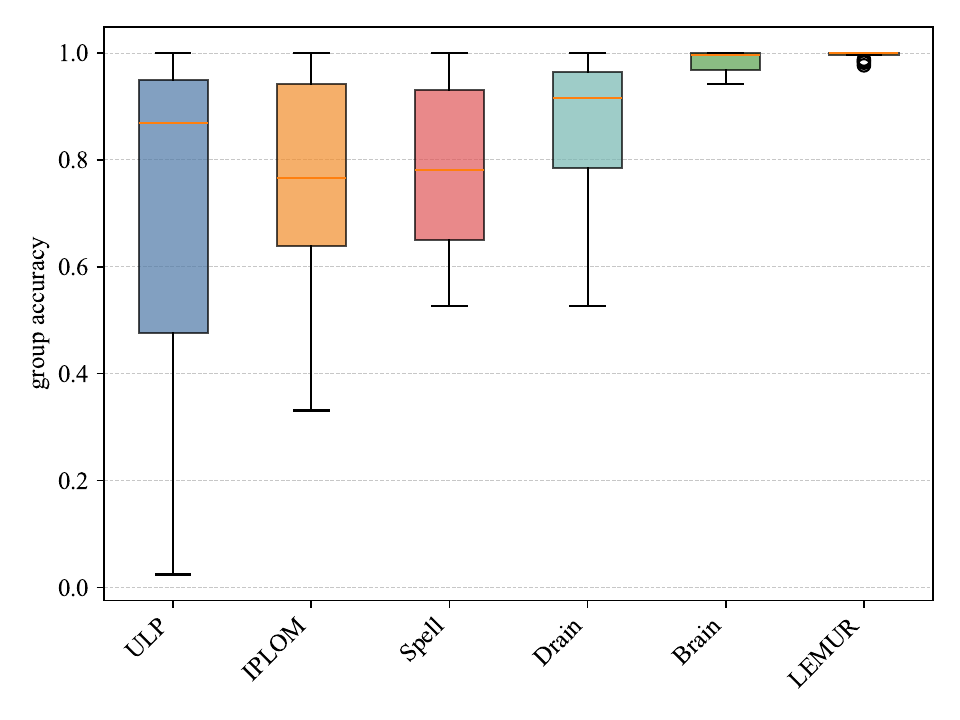}
        \caption{Boxplot of group accuracy on 16 benchmark datasets.}
        \label{fig:Boxplot_of_group_accuracy_on_16_benchmark_datasets}
    \end{subfigure}%
    \hfill
    \begin{subfigure}[t]{0.45\textwidth}
        \centering
        \includegraphics[width=\textwidth]{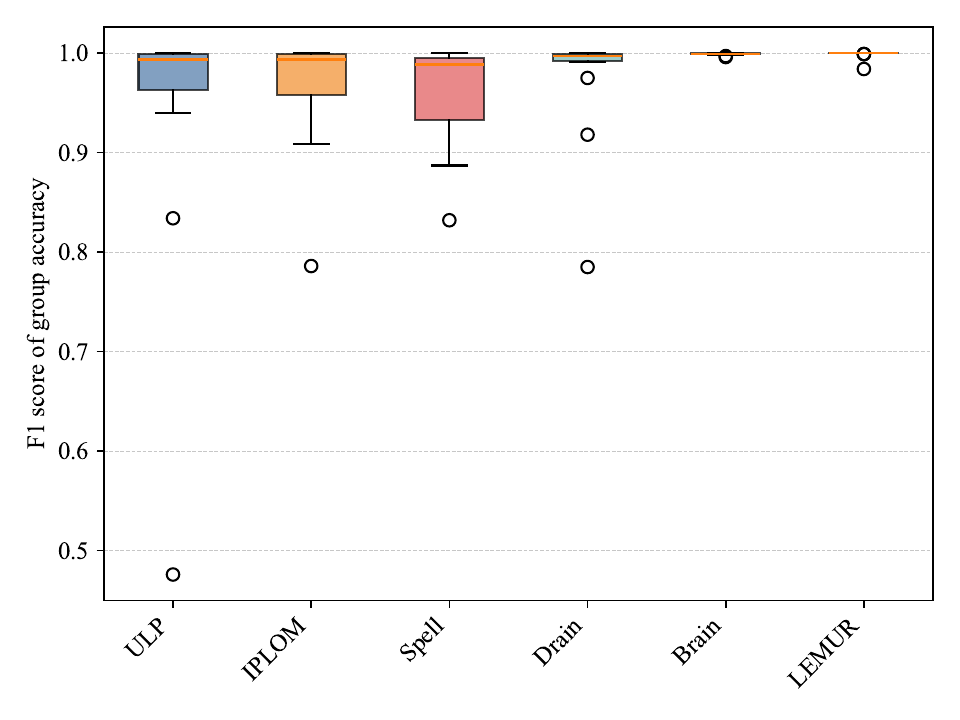}
        \caption{Boxplot of F1 score of group accuracy on 16 benchmark datasets.}
        \label{fig:Boxplot_of_F1_score_of_group_accuracy_on_16_benchmark_datasets}
    \end{subfigure}
    \caption{Comparison of group accuracy and F1 score on 16 benchmark datasets.}
    \label{fig:comparison}
    \vspace{-10pt}
\end{figure*}


\begin{figure*}[!t]
\includegraphics[width=1.0\linewidth]{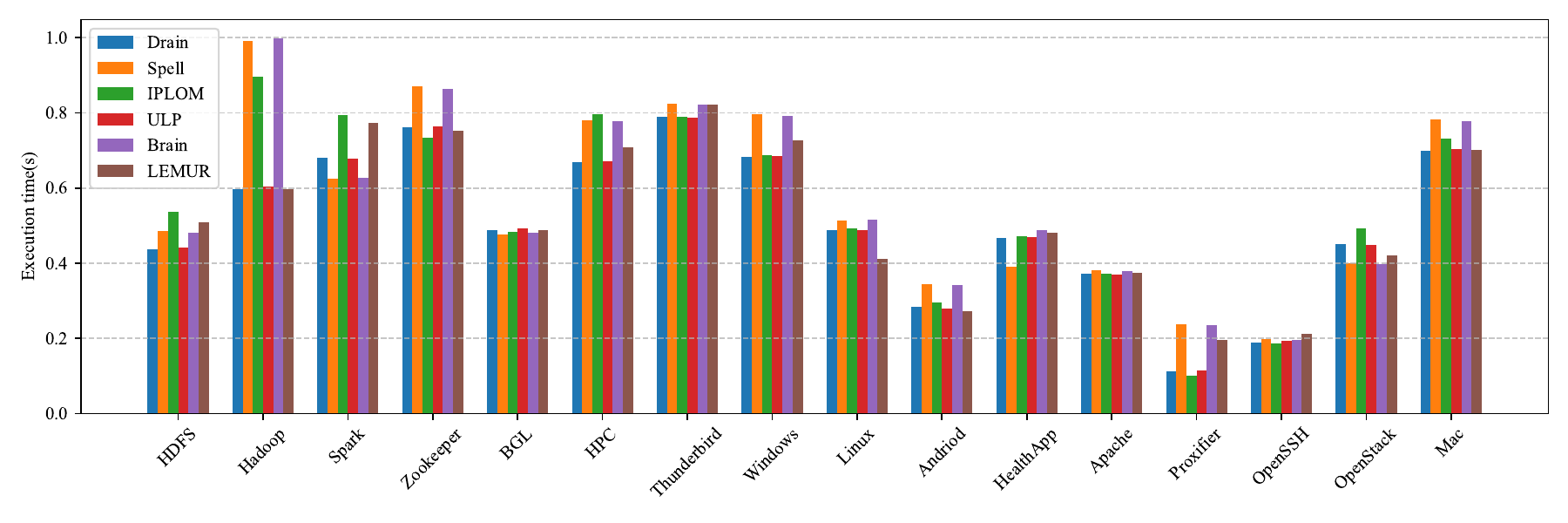}
\caption{Execution time for various datasets.}
\label{execution_time_for_various_datasets}
\vspace{-10pt}
\end{figure*}

\begin{figure*}[!t]
\includegraphics[width=1.0\textwidth]{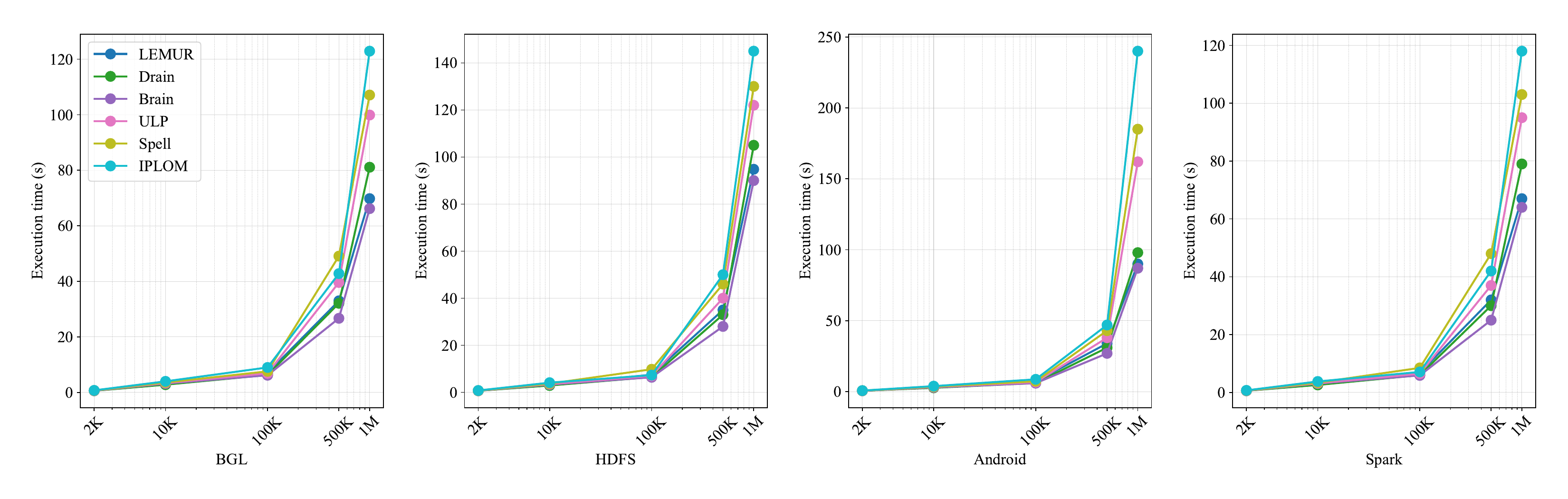}
\caption{Execution time for various dataset sizes.}
\label{execution_time_for_various_dataset_size}
\vspace{-10pt}
\end{figure*}

\begin{table}[!t]
\begin{center}
\resizebox{0.5\columnwidth}{!}{
\begin{tabular}{ccc}
\toprule
   Method         & Avg. FGA  & Avg. GA  \\
\midrule
   Random sampling & 0.843  &  0.718    \\
   First-token sampling & 0.913  &  0.804    \\
   Entropy sampling & 0.968  &  0.851    \\
   Entropy + First-token sampling (Ours) & 0.999 & 0.996 \\ 
\bottomrule
\end{tabular}
}
\end{center}
\label{sampling comparison}
\caption{Performance on different sampling methods.}
\vspace{-10pt}
\end{table}

\begin{table}[!t]
\begin{center}
\resizebox{\columnwidth}{!}{
\small
\begin{tabular}{lc|ccccccccccccccccc}
\toprule
\small Method & \small Metric & \small HDFS & \small Hadoop & \small Spark & \small Zookeeper & \small BGL & \small HPC & \small Thunderbird & \small Windows & \small Linux & \small Android & \small HealthApp & \small Apache & \small Proxifier & \small OpenSSH & \small OpenStack & \small Mac & \small \textbf{Average} \\
\midrule
\small \model{} w/o CoT & \small FGA
& 0.999 & 0.999 & \textbf{1} & 0.999 & 0.999 & 0.999 & \textbf{0.999} & \textbf{1} & \textbf{0.999} & 0.992 & \textbf{1} & \textbf{1} & \textbf{1} & \textbf{1} & \textbf{1} & 0.982 & 0.998 \\
\small \model{} w/o CoT & \small GA
& 0.998 & 0.977 & \textbf{1} & 0.995 & 0.989 & 0.996 & 0.959 & \textbf{1} & 0.986 & 0.956 & \textbf{1} & \textbf{1} & \textbf{1} & \textbf{1} & \textbf{1} & 0.956 & 0.988 \\
\small \model{} & \small FGA
& \textbf{1} & \textbf{1} & \textbf{1} & \textbf{1} & \textbf{1} & \textbf{1} & \textbf{0.999} & \textbf{1} & \textbf{0.999} & \textbf{0.999} & \textbf{1} & \textbf{1} & \textbf{1} & \textbf{1} & \textbf{1} & \textbf{0.984} & \textbf{0.999} \\
\small \model{} & \small GA
& \textbf{1} & \textbf{1} & \textbf{1} & \textbf{1} & \textbf{1} & \textbf{1} & \textbf{0.982} & \textbf{1} & \textbf{0.988} & \textbf{0.995} & \textbf{1} & \textbf{1} & \textbf{1} & \textbf{1} & \textbf{1} & \textbf{0.977} & \textbf{0.996} \\
\bottomrule
\end{tabular}
}
\end{center}
\caption{FGA and GA on LogHub between \model{} with and without CoT.}
\label{FGA_and_GA_on_LogHub_between_model_and_model_without_cot}
\vspace{-10pt}
\end{table}

\begin{table}[!t]
\begin{center}
\resizebox{0.5\columnwidth}{!}{
\begin{tabular}{lcc}
\toprule
Configuration & Avg.FGA & Avg.GA \\
\midrule
Default (Our Method) & 0.999 & 0.996 \\
KMeans Clustering \citep{sinaga2020unsupervised} & 0.685 & 0.698 \\
DBSCAN Clustering \citep{khan2014dbscan} & 0.731 & 0.744 \\
Our Method w/o Bucket Grouping & 0.355 & 0.394 \\
\bottomrule
\end{tabular}
}
\end{center}
\label{tab:additional_ablation_results}
\caption{Additional Ablation Studies on \model{}}
\vspace{-10pt}
\end{table}

In Table \ref{public_data_accuracy}, \model{}, an unsupervised LLM-related model, exhibits remarkable effectiveness and performance in the field of log parsing. When compared to Brain, the current state-of-the-art in unsupervised models, \model{} demonstrates superior or comparable results across various datasets, underscoring its efficacy in unsupervised learning without the need for annotated data. Moreover, even when juxtaposed with supervised models like LogPPT and LLMParser, \model{} shows equal or better performance in most datasets, based on FGA and GA metrics. This is particularly noteworthy as it highlights the capability of \model{} to match or exceed the performance levels of supervised models, despite the absence of explicit label guidance.

In Figure~\ref{fig:comparison}, \model{} shows the robust performance in GA and FGA. The consistency and robustness of \model{} are evident in its high FGA and GA across diverse datasets such as HDFS, Hadoop, Spark, etc. This consistency emphasizes its adaptability and robustness to various log parsing requirements. In conclusion, as an unsupervised LLM-related model, \model{} significantly stands out in the log parsing domain. 

Figure~\ref{execution_time_for_various_datasets} presents a comprehensive analysis of the execution times for multiple algorithms: \model{}, Brain, Drain, IPLOM, Spell, and ULP, across various datasets, comprising 2K data points each. This figure provides critical insights into the efficiency and scalability of these algorithms in processing data. Notably, the execution time of \model{} stands out for its relative brevity across the majority of the datasets. LEMUR demonstrates a significantly reduced execution time on several key datasets, including Hadoop, BGL, Windows, Andriod, and Mac, highlighting its efficiency and optimized processing capabilities. Furthermore, we extend this analysis to evaluate the performance scalability of these algorithms across four distinct datasets: BGL, HDFS, Android, and Spark. This extension, as depicted in Figure~\ref{execution_time_for_various_dataset_size}, encompasses varying dataset sizes, thereby providing a more nuanced understanding of each algorithm's adaptability and performance under different data volume conditions in real-world scenarios where dataset sizes can vary significantly. 

Compared to all unsupervised methods, we have achieved more superior performance in the preliminary stage before employing LLM. Furthermore, compared to other LLM-based methods, which are mostly supervised, such as LogPPT and LLMParser, which are essentially inference after fine-tuning, the training process consumes more GPU resources for local deployment or involves larger data transmission over the network for closed-source ChatGPT. LEMUR achieves better results without fine-tuning the LLM. Whether for local deployment inference or remote API, it only utilizes GPU resources for inference, not for fine-tuning. Thus, it requires fewer GPU resources or less network latency in data transmission. In summary, LEMUR has achieved superior performance and unique advantages for unsupervised log parsing or supervised fine-tuning of LLM.

\section{Ablation}
\subsection{Effect of Entropy Sampling}
In our comparative analysis, as delineated in Table 4, we meticulously evaluate four distinct sampling methodologies: Random sampling, First-token sampling, Entropy sampling, and a hybrid approach combining Entropy and First-token sampling. Our results, derived from a systematic and empirical evaluation, reveal that the hybrid Entropy + First-token sampling method exhibits superior performance over the other techniques under consideration. The enhanced effectiveness of this method is likely attributable to its synergistic integration of the entropy-based selection criterion, which effectively captures the variability within the data, and the first-token approach, which ensures the representativeness of the initial elements. This fusion of strategies enables a more nuanced and effective sampling process, as evidenced by our empirical findings.

As illustrated in Table 5, which provides a detailed comparison between two versions of \model{}: one implementing the three-hop Chain-of-Thought approach and the other without it. The left columns of the table present the performance metrics of \model{} devoid of Chain-of-Thought (CoT), while the right columns display the outcomes following the integration of the CoT methodology. The data delineates a useful, albeit modest, improvement in the performance of \model{} when augmented with the CoT approach. This enhancement is particularly evident in the FGA and GA metrics across a range of datasets, including Hadoop, BGL, HPC, Thunderbird, Linux, Android, and Mac.

The observed limited degree of enhancement can be ascribed to two primary factors. The first is the already high baseline performance of \model{} without the incorporation of CoT, which inherently constrains the potential for significant further improvements. The second factor pertains to the relative simplicity of the loghub datasets. In these scenarios, the application of a sophisticated methodology like CoT results in only marginal improvements, primarily because the datasets do not present sufficient complexity to fully exploit and showcase the enhanced capabilities of the CoT-augmented \model{}.

\subsection{Additional Ablation Studies on \model{}}

To further validate the robustness of \model{}, we conducted ablation studies focusing on the impact of different clustering algorithms and the effectiveness of the bucket grouping strategy. Specifically, we experimented with replacing our clustering mechanism with KMeans \citep{sinaga2020unsupervised} and DBSCAN \citep{khan2014dbscan} and evaluated the performance when disabling the bucket grouping based on the length assumption. These modifications aim to probe the sensitivity of \model{} to these components.

The results, as illustrated in Table 6, highlight the critical role of the selected clustering algorithm and bucket grouping strategy in \model{}'s performance. The substantial decrease in Avg.FGA and Avg.GA when employing alternative clustering methods or omitting the bucket grouping underscores their importance in achieving optimal outcomes. This ablation study confirms the necessity of careful component selection and validates the robustness of \model{}'s methodology.

\section{Related Work}
\subsection{Log Parser}
In the evolving field of automatic log analysis, crucial for distributed systems and cloud computing, significant progress has been made in log parsing techniques, categorized into frequent pattern mining, clustering, and heuristics rules. Frequent pattern mining is exemplified by SLCT \citep{slct} which groups logs based on token frequency, and LogCluster \citep{logcluster} which removes positional constraints in log grouping. Clustering approaches include LogMine \citep{logmine} with its multi-layered clustering system, LKE\citep{lke} using edit distance and position weighing, SHISO \citep{shiso} improving efficiency through hierarchical clustering, LenMa \citep{lenma} employing token length vectors, and LPV \citep{lpv} which uses semantic vectors from word2vec. In the heuristics rules category, IPLOM \citep{iplom} partitions logs by length and token position, Spell \citep{spell} approaches parsing as the longest common sequential problem, Drain \citep{drain} groups logs by length and prefixes for template updates, and Prefix-Graph \citep{prefix_graph} merges prefix trees into graphs for template generation. Recent advancements have introduced deep learning-based algorithms like Nulog \citep{nulog} Uniparser \citep{uniparser}, and LogAP \citep{logap} utilizing comparative learning and machine translation for parsing. However, these deep learning methods face challenges in efficiency and high operational costs due to GPU requirements.

\subsection{Large Language Model}
Language modeling using self-supervised learning and large-scale data, significantly enhances various natural language processing tasks. Specifically, pre-training a Transformer decoder \citep{gpt4,transformer,codearena,xcoder,execrepobench,ouyang2022training,wei2022finetuned,fingpt,lawgpt,mdeval,mceval,zhang2024mabcmultiagentblockchaininspiredcollaboration,zhang2024eclipsesemanticentropylcscrosslingual,10888405} aids in unconditional text generation. Performance improvements~\citep{bertsum,alm,hlt_mt,um4,xmt,microsoft_wmt2021,soft_template} in diverse tasks have been linked to the enlargement of Pre-training Language Models (PLMs) by increasing model or data size. This has led to the creation of increasingly larger PLMs, such as GPT-3 with 175 billion parameters and PaLM with 540 billion \citep{palm2}, guided by the scaling laws of large language models \citep{scaling_laws_gmlm}. Despite their similar architectures and pre-training tasks, larger PLMs, such as GPT-4 \citep{gpt4}, exhibit unique behaviors and emergent abilities, excelling in complex tasks. A prime example is ChatGPT, adapting GPT-series LLMs for engaging dialogues, and showcasing advanced conversational skills. Fine-tuning LLMs on various datasets \citep{CoT} yields promising results, using human or LLM-created prompts for instruction tuning and refining generations. Chain-of-thought prompting \citep{CoT}, where models explain their reasoning for complex problems, and RLHF \citep{rlhf}, a strategy significantly enhance their performance.

\section{Conclusion}
To enhance log analysis in complex software systems, we propose \model{}. This framework replaces traditional rule-based methods with an information entropy-based sampling for efficient log clustering and employs large language models (LLMs) for advanced semantic comprehension in template merging. Information entropy streamlines the process of distinguishing between disparate log messages based on their inherent informational content for efficient log clustering. \model{} has demonstrated superior performance and efficiency in log parsing, validated through extensive tests on large-scale datasets.



\bibliography{iclr2025_conference}
\bibliographystyle{iclr2025_conference}


\end{document}